\documentstyle[11pt]{article}
\newcommand{\EQ}{\begin{equation}}
\newcommand{\EN}{\end{equation}}
\begin{document}
\pagestyle{empty}
\def\singlespacing{\baselineskip=12pt}
\def\doublespacing{\baselineskip=24pt}
\doublespacing


\bigskip

\begin{large}
\centerline{Distribution of domain sizes in the zero temperature }
\centerline{Glauber dynamics of the 1 D Potts model}
\end{large}

\begin{center}
\bigskip
{Bernard Derrida$^{1,2}$,
 and Reuven Zeitak$^{1}$}
\end{center}
\bigskip
{\small
$^1$Laboratoire de Physique Statistique$^{3}$, ENS,
 24 rue Lhomond, F-75231 Paris cedex 05, France\\
$^2$Service de Physique Th\'{e}orique,
CE Saclay, F-91191 Gif sur Yvette, France}
\addtocounter{footnote}{3}
\footnotetext{ Laboratoire associ\'e aux Universit\'es
Pierre et Marie Curie, Denis Diderot et au CNRS }

\bigskip
\bigskip

\begin{center}
{\bf ABSTRACT}
\end{center}

\medskip

For the zero temperature Glauber dynamics of the $q$-state Potts model,
we calculate the exact  distribution of domain sizes
by mapping the problem on an exactly soluble
 one-species coagulation model ($A+A\rightarrow
A$). In the long time limit, this distribution is universal
and from its (complicated) exact expression, we extract its behavior
in various regimes. Our results are tested in a simulation and compared
to the predictions of a simple approximation proposed recently.
Considering the dynamics of domain walls as a reaction diffusion model
$ A+A \to \ A $ with probability $ (q-2)/(q-1)$ and
$A+A \to \emptyset $ with probability  $1/(q-1)$, we calculate the pair
correlation
function in the long time regime.

\bigskip
\bigskip
\bigskip
\bigskip
\bigskip
\bigskip
\noindent PACS:  {02.50.+s, 05.40.+j, 82.20-w} \\
Short title: Distribution of domain sizes \\

\newpage
\pagestyle{plain}

\section{Introduction}
When a ferromagnetic system is quenched from the high temperature phase to
a temperature below its Curie temperature,
one observes a pattern of growing domains.
In the long time limit,  when the typical size of  domains
becomes much larger than the lattice spacing (or the correlation length)
but is still small compared to the system size, the domains form a
(statistically) self similar  structure (see \cite{Bray} and references
therein). For a non-conserved order parameter,
 it is well established that the size of the domains grows  with time $t$ like
$t^{1/2}$.
Much less is known on the distribution of domain sizes.

The purpose of  the present paper is to give the exact distribution
of domain sizes in the case of the  one dimensional $q$-state Potts model
evolving according to
 zero temperature  Glauber dynamics \cite{Glauber} (in $1d$, the low
temperature phase reduces  only to
zero temperature).
As \cite{Br,AF} the average domain size grows with time like $t^{1/2}$, the
distribution
of domain sizes  can only be determined up to a change of scale
and we will  rescale the  length $x$  of the domains so that for  the
distribution  $g(x;q)$,  the average domain size is unity
($\int x  \ g(x;q) \ dx= 1$).  So far, this distribution has been calculated
only for $q=\infty$  \cite{ABD,DGY}  because it is related to the probability
that two walkers do not meet up to time $t$.
Our goal, here, is to extend this result to arbitrary $q>1$.

Our approach is an generalization of a calculation
done recently \cite{DHP,DHP1} to obtain the fraction of persistent spins (i.e.
spins which never flip up to time $t$).  We  calculate
the probability $Q(N)$ that $N$ given  consecutive  spins are parallel
at time $t$ and explain how $g(x;q)$ can be extracted from $Q(N)$.
As for the number of persistent spins, the full expression
 of $Q(N)$
is rather complicated but   it can be used to write explicit  formulae
in various limits.

It has been noticed for a long time that the zero temperature Glauber
dynamics  of the $1d$ Potts model  is fully equivalent to a
single species  reaction-diffusion model \cite{AF,ABD,DGY,Ra,Pr}. If one
represents each
domain wall by a particle $A$ and if the initial spin configuration
is random with no correlation, it is easy to show that
the particles $A$ diffuse along the line and that whenever 2 particles
sit  on the same bond, they instantaneously react according
to
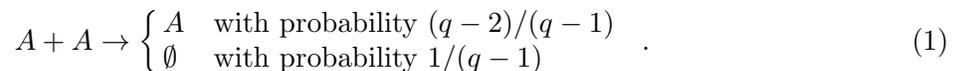
\begin{equation}
 A+A \to \cases{ A & \makebox{with probability} $ (q-2)/(q-1)$ \cr
 \emptyset & \makebox{with probability}  $1/(q-1)$ \cr}  \ .
\label{reactdif}
\end{equation}
We will see
that  several of our results   can be reinterpreted
as properties of this $1d$ reaction diffusion problem.

An easy way to implement the zero  temperature dynamics  of the $1d$
$q$-state Potts model is to say that during every infinitesimal time interval
$dt$, each spin $S_i(t)$ is updated according to
$$S_i(t+dt) = \cases{ S_i(t) & with probability $1 -2 dt$ \cr
S_{i-1}(t) & with probability $ dt$ \cr
S_{i+1}(t) & with probability $ dt$ \cr} \ .
$$
This shows  the close  analogy with random walk
problems \cite{voter}. As in \cite{DHP,DHP1,Der}, this analogy will be the
basis of our calculation.

The paper is organized as follows. In section 2, we  recall
several   properties of random walks which are used in
the following sections and the relation between random walks and the
zero temperature  Glauber dynamics of a spin chain. In section 3, we  obtain,
 using
several relations derived in \cite{DHP1}, the probability $Q(N)$ that, at
 time $t$, $N$ given  consecutive spins are parallel and we show how the
 distribution of domain sizes follows from the knowledge of the $Q(N)$.
 In section 4, we
give more explicit expressions in various limits  and in section 5 we
 compare our predictions with the results of simulations.
In section 6, we calculate the pair correlation for the reaction diffusion
model (\ref{reactdif}).

\section{Properties of random walks in one dimension and the $q=\infty$ case}
Let us first recall some properties of random walks in
 one dimension. Several
of these properties are well known  (in particular they can be obtained
by representing   the walkers  as free fermions \cite{dG1,dG2,dG3}) but
we spend some time discussing them because they are
essential to the understanding of the following sections. These properties will
include
the probability $c_{i,j}$ that two walkers starting at positions $x_i < x_j$ do
not meet up
to time $t$, the probability $c_{1,2...,2n}^{(n)}$ that no pair meets up to
time $t$ between
$2n$ walkers starting at positions $x_1 < x_2 <.. x_{2n}$.

Consider a random walker on a $1d$ lattice which hops, during each
infinitesimal
time interval to its right with probability $dt$, to its left with probability
$dt$ (and of course remains at the same position with probability $1 - 2 dt$).
The probability $p_t(y,x)$ of finding the walker at position
$y$ at time $t$ given that it was initially (at time $0$) at position $x$
evolves
according to
\begin{equation}
{d \over dt} p_t(y,x) = p_t(y+1,x) + p_t(y-1,x) -2 p_t(y,x) =
 p_t(y,x+1) + p_t(y,x-1) -2 p_t(y,x)
\end{equation}
and the solution is
\begin{equation}
 p_t(y,x) ={1 \over 2 \pi} \int_{-\pi}^\pi d \theta
\ \cos(x-y) \theta \
 \  e^{-2(1 - \cos \theta)t}   \ .
\end{equation}
Consider now 2 walkers on this $1d$ lattice starting at positions
$x_i$ and $x_j$ with $x_i < x_j$. The probability   $C_t(x_i,x_j) \equiv
c_{i,j}$
 that these two walkers never meet up to time $t$
evolves according to
\begin{equation}
{d \over dt} C_t(x,x') =C_t(x+1,x')+C_t(x-1,x')+C_t(x,x'+1)+C_t(x,x'-1) -4 \
C_t(x,x')
\label{evo}
\end{equation}
with the boundary conditions that
$C_t(x,x)=0$  at any time $t$ and that
$C_0(x,x')=1$ if $x<x'$.
It is easy to check that the following expression
\begin{equation}
c_{i,j} = C_t(x_i,x_j) =
{1 \over 2 \pi} \int_{-\pi}^\pi d \theta
  { \sin \theta \sin(x_j-x_i) \theta \over 1 - \cos \theta}
 \  e^{-4(1 - \cos \theta)t}
\label{cij}
\end{equation}
satisfies (\ref{evo}).
It is  remarkable \cite{DHP1} that the probabilities of all the meeting
events of $N$  coalescing random walkers starting at positions $x_1 \leq  x_2
\leq .. \leq  x_N$
can be expressed in terms of the matrix $c_{i,j}$. For example, for 3 walkers
starting at positions $x_1 \leq x_2 \leq x_3$, the probability that none of
them  meets  any of the other two  up to time $t$ is given by
$$c_{1,2}+c_{2,3} - c_{1,3} \ .$$ For 4 walkers starting at $x_1 \leq x_2 \leq
x_3 \leq x_4$, the probability that no pair meets up to time $t$ is given by
\begin{equation}
c^{(2)}_{1,2,3,4} = c_{1,2} \ c_{3,4} \ + \ c_{1,4} \ c_{2,3} \ - \ c_{1,3} \
c_{2,4} \ .
\label{PF2}
\end{equation}
Below, we will use  the fact  \cite{DHP1} that  the probability that, up to
time $t$,  no pair meets
among $2n$ walkers starting  at positions $x_1 \leq x_2 \leq ... x_{2n}$, is
given by the Pfaffian of the $2n \times 2n$ matrix $c_{i,j}$
\begin{equation}
c^{(n)}_{1,2,...,{2n-1},{2n}}=\frac{1}{2^n \ n!}\sum_{\sigma}
 \ \epsilon(\sigma) \  c_{{\sigma(1)},{\sigma(2)}} \cdots
c_{{\sigma(2n-1)},{\sigma(2n)}}
\label{PFN}
\end{equation}
where the sum runs over all the permutations  $\sigma$ of the
indices $\{1,2,\cdots,{2n}\}$,
$\epsilon(\sigma)$ is the signature
of the permutation $\sigma$   and  the matrix $c_{i,j}$ is  antisymmetrized
$$c_{i,j}= -  \ c_{j,i} \ < 0  \ \ \ \makebox{when} \ \ \  i > j \ . $$

There are several ways of deriving (\ref{PF2},\ref{PFN}), for example by
considering the walkers as free fermions \cite{dG1,dG2,dG3,DHP1} or by using
the method of images.
One can also simply write
the equations similar to (\ref{evo}) which govern the evolution of these
probabilities and
check that (\ref{PF2},\ref{PFN}) do satisfy these equations with
the right boundary conditions when $c_{i,j}$ is solution of  (\ref{evo}).
(note that the above relations would not be valid in dimension higher than 1:
it is only  in $1d$ that whenever 1 and 2 do not meet, it  implies that 1 and 3
do not meet because of the order of the walkers along the line).

Apart the probabilities that the walkers never meet, the knowledge of the
matrix $c_{i,j}$ allows one to calculate all the meeting probabilities between
coalescing random walkers \cite{DHP1}. For example, when four walkers start
 at positions $x_1 < x_2< x_3< x_4$, one finds for the probability
$\psi_{12,34}$ that $1$ and $2$ coalesce, $3$ and $4$ coalesce but $2$ and $3$
do not coalesce before time $t$
\begin{equation}
\psi_{12,34} =
  c_{1,3}  +c_{2,4} +c^{(2)}_{1,2,3,4} -c_{1,2} -c_{2,3} - c_{3, 4}  \
\label{1234}
\end{equation}
and for the probability $\psi_{1,23,4}+ \psi_{1,2,3,4}$ that  $1,2$ do not meet
and
$3,4$ do not meet (without saying whether $2$ and $3$ meet or not)
\begin{equation}
\psi_{1,23,4} +
\psi_{1,2,3,4} =
  c_{1,3}  +c_{2,4} +c^{(2)}_{1,2,3,4} -c_{1,4} -c_{2,3}   \  .
\label{1234a}
\end{equation}

In the long time limit, all the above expressions
(\ref{PF2},\ref{PFN},\ref{1234},\ref{1234a}) remain valid with $c_{i,j}$
replaced by the asymptotic expression of
(\ref{cij})
\begin{equation}
c_{i,j} = C_t(x_i,x_j) =  f \left( {x_j - x_i \over   \sqrt{8t}} \right)
\label{cijbis}
\end{equation}
\begin{equation}
f(z)=
{2 \over \sqrt{\pi}} \int_0^{z   } e^{-u^2} \ du \ .
\label{deff}
\end{equation}
\subsection{Relation between random walks and the dynamics of the spin chain:}
As in \cite{Der,DHP,DHP1}, we are going to use the close analogy between
the problem of   coalescing random walkers
and the properties of a spin chain evolving according to zero temperature
dynamics:
assume that we want to calculate, for a  Potts chain  with no correlation  in
the initial condition, the probability that spins at position $x_1< x_2< ... <
x_N$ are equal at time $t$.
One can  consider $N$ coalescing random walkers starting at positions
$x_1<x_2<... <x_N$ and if $P(m,t)$ denotes the probability that at time $t$,
there are $m$ walkers left in the system,
one has \cite{DHP,DHP1}
\begin{equation}
\makebox{Probability}\{S_{x_1}(t)= S_{x_2}(t)=... S_{x_N}(t) \} = \sum_{m=1}^N
P(m,t)  {1 \over q^{m-1}} \ .
\label{Pm}
\end{equation}
This expression can be understood by noticing  that initially the spins are
random and
that the probability that $m$ spins have the same color in the initial
condition is just
$1/q^{m-1}$.
\subsection{ The case $q=\infty$:}
In the limit $q \to \infty$ only the first term of the sum contributes and
one finds that
\begin{equation}
\makebox{Probability} \{ S_{x_1}(t)= S_{x_2}(t)=... S_{x_N}(t) \} = 1 -
C_t(x_1,x_N) \ .
\label{123N}
\end{equation}
This expression can easily be understood: for $q= \infty$, all the spins in the
initial configuration are different; if
$S_{x_1}(t)= S_{x_N}(t)$, this means that the two walkers starting
at $x_1$ and $x_N$ have met before time $t$ and therefore, at time $t$, all the
  spins located
between $x_1$ and $x_N$ are equal.
It can be used to calculate the distribution of domain sizes  in the $q \to
\infty$ limit: if $p_1(l)$ is the   density of domains of length $l$, one has
\begin{equation}
\makebox{Probability}\{S_{x_1}(t)= S_{x_2}(t)=... S_{x_N}(t) \} =  \sum_{l \geq
x_N-x_1} (l-x_N+x_1) p_1(l)
 \ .
\label{p1}
\end{equation}
This leads to
\begin{equation}
p_1(l)=2 C_t(0,l)- C_t(0,l+1) - C_t(0,l-1)
\end{equation}
and in the long time limit
\begin{equation}
p_1(l)=  - {1 \over 8 t} f'' \left( {l \over  \sqrt{8t}} \right)
\end{equation}
For $q= \infty$, this allows one to calculate the average domain size at time
$t$
\begin{equation}
\langle l \rangle_\infty = {\sum_l l p_1(l)   \over \sum_l p_1(l)} \simeq {
\sqrt{8t} \over f'(0)} =
 {  \sqrt{8t} \over f'(0)} =  \sqrt{2 \pi t}
\label{la}
\end{equation}
and the distribution  $g(x;\infty)$ of domain sizes \cite{DGY}  (normalized
such that the average  size is $1$)
\begin{equation}
g(x;\infty) =  \langle l \rangle \ { p_1( \langle l \rangle x) \over \sum_{l'}
p_1(l')} = {\pi \over 2} \  x \  e^{-x^2 \pi /4}  \ .
\label{ginf}
\end{equation}
For similar reasons, it is easy to see that for $q=\infty$, the probability
that $S_{x_1}(t)= S_{x_2}(t) \neq S_{x_3}(t) = S_{x_4}(t) $ is given
by
\begin{equation}
\makebox{Probability} \{
S_{x_1}(t)= S_{x_2}(t) \neq S_{x_3}(t) = S_{x_4}(t) \}
= \psi_{12,34}
\end{equation}
This can be used to obtain
$p_2(l_1,l_2)$,  the density of domains of  length $l_1$ followed by a domain
of length $l_2$ (because for $q=\infty$, whenever two spins have a certain
color, all the spins between them  have the same color)
\begin{equation}
\makebox{Probability} \{
S_{x_1}(t)= S_{x_2}(t) \neq S_{x_2+1}(t) = S_{x_4}(t) \} =
\sum_{l_1 \geq x_2-x_1} \ \
\sum_{l_2 \geq x_4-x_2-1} p_2(l_1,l_2) \ .
\end{equation}
and consequently (\ref{1234}-\ref{deff}), the (normalized)  distribution
$g_2(x,y;\infty)$ of  neighboring domain sizes is given by
\begin{equation}
g_2(x,y;\infty) =
  \langle l \rangle^2 \ { p_2( \langle l \rangle x, \langle l \rangle y) \over
\sum_{l'} p_1(l')} = {\pi \over 2} (x+y)  \left[e^{-(x^2+ y^2) \pi /4}  -
e^{-(x+y)^2 \pi /4} \right]
\label{g2inf}
\end{equation}
This means that the lengths
 of consecutive domains in the limit $q \to \infty$
are correlated. In particular, if $l_1$ and $l_2$ are the lengths of two
consecutive domains, one has
\begin{equation}
{\langle l_1 \l_2 \rangle \over \langle l \rangle^2} = \int_0^\infty dx
\int_0^\infty dy  \ g_2(x,y;\infty) \ x \ y =  {3 \over \pi} \neq 1
\label{COR}
\end{equation}
 \\ {\bf Remark 1:}
The knowledge of  the distribution of domain sizes  and of all the correlations
between consecutive domain sizes for $q=\infty$  gives the distribution of
domain sizes for arbitrary $q$. If $l_1,l_2,....l_n,...$ are the lengths of
consecutive domains when $q=\infty$, one can obtain the length $l$ of a typical
domain for finite $q$ by
\begin{equation}
l = \cases{ l_1  & with probability ${q-1 \over q}$ \cr
l_1 + l_2  & with probability ${q-1 \over q^2}$ \cr
...  &  \cr
l_1 + l_2 + ... l_n  & with probability ${q-1 \over q^n}$ \cr
...  &  \cr}
\label{fq}
\end{equation}
This is because  one way of performing the dynamics at finite $q$ is to first
make the system evolve with all the spins different in the initial condition
(as in the case $q=\infty$) and then  suppress the domain walls with
probability $1/q$. This is due to the fact that for  finite $q$, two initial
values are identical with probability $1/q$.

A consequence of (\ref{fq}) is that, in the long time limit,  the average
length for finite $q$ is given
by (\ref{la})
\begin{equation}
\langle l \rangle_q  =
{q \over q-1} \langle l \rangle_\infty
  \simeq {q \over q-1}
  \sqrt{ 2\pi t}
\label{la1}
\end{equation}
It is clear also from (\ref{fq}) that the calculation of any other moment
$\langle l^n \rangle$ of  the length  would require the knowledge of all the
correlations between the  $l_i$.
 \\ {\bf Remark 2:}
If  these correlations between  successive domain sizes for $q=\infty$ were
absent, the calculation of the normalized distribution $g(x;q)$ for arbitrary
$q$  would be straightforward: neglecting the correlations in (\ref{fq}) would
give for the generating function of $g(x;q)$
\begin{equation}
\int_0^\infty  dx g(x;q) e^{\alpha q x /(q-1)} dx =
{ (q-1)\int_0^\infty dx g(x;\infty) e^{\alpha  x } dx
 \over   q - \int_0^\infty dx g(x;\infty) e^{\alpha  x } dx  }
\label{indep}
\end{equation}
Then using the expression (\ref{ginf}), one would find (by analyzing the limit
$\alpha \to -\infty$)
\begin{equation}
g(x;q)= {\pi \over 2} {q \over q-1} x - {\pi^2 \over 24} {(3q-1)q^2 \over
(q-1)^3}  x^3 +
 {\pi^3 \over 960} {(15q^2-6q+1)q^3 \over (q-1)^5}  x^5+ O(x^7)
\label{ind1}
\end{equation}
(and  by analyzing the pole in $\alpha$)
\begin{equation}
g(x;q) \simeq \exp \left[-A(q)x+B(q) \right]
 \ \ \  \makebox{for large} \ x
\label{xlarge}
\end{equation}
where $A(q)$  is the root of
\begin{equation}
{\pi \over 2} \int_0^\infty  dx \ x \exp \left[ - {\pi x^2 \over 4} + {(q-1)x
\over q} A(q) \right] = q
\label{ind2}
\end{equation}
and $B(q)$ is given by
\begin{equation}
B(q) =
 \log \left\{ {q^2  \pi A(q)   \over q \pi + 2 (q-1) A(q)^2 }\right\} \ .
\label{ind3}
\end{equation}
For $q=2$, this is precisely the prediction  recently given in \cite{ALEMA},
based on the assumption that the intervals $l_1,l_2,..$ are uncorrelated. We
have already seen
(\ref{g2inf},\ref{COR}) that this assumption is not valid. Our goal in  the
following
sections is to obtain the true distribution $g(x;q)$ where these correlations
have been taken into account. We will see that the exact expression of $g(x;q)$
disagrees with (\ref{ind1},\ref{ind2},\ref{ind3}).
\section{Arbitrary correlation functions}
Our solution for the distribution of size of domains is  based on our ability
to write exact expressions
 for all correlation functions (valid at an arbitrary time)
when the initial condition is random with no correlation (i.e. each spin takes
initially one of the $q$ possible colors with equal probability).
These exact expressions involve the probabilities $c^{(n)}$ given by
(\ref{cij},\ref{PF2},\ref{PFN}). One can show in particular that
the probability that   $N$ spins located
at positions $x_1 \leq x_2 \leq .. x_N$  are identical at time $t$ is given by
\begin{eqnarray}
\makebox{Prob} \{ S_{x_1}(t)= ... S_{x_N}(t)\}  = 1
- \mu \sum_{i=1}^{N-1} c_{i,i+1} + \mu^2 \sum_{i < j} c_{i,i+1,j,j+1}^{(2)}
- \mu^3 \sum_{i < j < k } c_{i,i+1,j,j+1,k,k+1}^{(3)} +\cdots
\nonumber \\
-  \lambda \left \{\mu  c_{1,N} -  \mu^2 \sum_{i } c_{1,i,i+1,N}^{(2)}
+  \mu^3 \sum_{i < j  } c_{1,i,i+1,j,j+1,N}^{(3)} - \cdots  \right\}
\label{ProN}
\end{eqnarray}
where
\begin{equation}
\lambda = q-1
\label{LAMBDA}
\end{equation}
\begin{equation}
\mu = {q-1 \over q^2} \ .
\label{MU}
\end{equation}
(note that as long as $N$ is finite, (\ref{ProN})  is the sum of a finite
number of terms and is therefore
a polynomial in the variable $\mu$).

The proof of (\ref{ProN}) is exactly the same as the one given in \cite{DHP}:
one  shows that (\ref{ProN}) is equivalent to
  (\ref{Pm}) for  $N$ coalescing random walkers starting at positions
$x_1 \leq x_2 \leq ... x_N$
by considering all the possible coalescing events between the $N$ walkers.
The weight $a_m$  in (\ref{ProN})  of an event
where at time $t$ the $N$ walkers have merged into $m$ walkers
is given by
\begin{eqnarray}
a_m=
1 - (m-1) \mu + {(m-2)(m-3) \over 2!} \mu^2  - {(m-3)(m-4)(m-5) \over 3!} \mu^3
 \cdots
\nonumber \\
- \lambda \left(
 \mu - {(m-3) } \mu^2  + {(m-4)(m-5) \over 2!} \mu^3  \cdots \right)
\label{am1}
\end{eqnarray}
which is  just $a_m = 1/q^{m-1}$  (see (\ref{LAMBDA}) and (\ref{MU})).
This therefore proves that (\ref{ProN}) is equivalent to (\ref{Pm}).

In \cite{DHP1},  several ways of rewriting (\ref{ProN}) were given.
In particular it was shown that (\ref{ProN}) can be rewritten as the square
root
of a determinant
\begin{eqnarray}
\makebox{Prob} \{ S_{x_1}(t) = ...S_{x_N}(t) \}
= \left| \left(\begin{array}{cccccc}
                  1&0&0&0&& \\
                  0 &1&0&0&& \\
                  0 &0&.&&& \\
                   &&&.&& \\
                   &&&&.& \\
                   &&&&&1 \\
                   \end{array}
                 \right)
+ \mu
 \left(\begin{array}{cccccc}
                  0&1&0&&0& \lambda\\
                  -1 &0&1&0&&0 \\
                  0 &-1&0&1&0& \\
                   &&.&&.& \\
                  0 &&&.&&1 \\
                  -\lambda &0&&&-1&0 \\
                   \end{array}
                 \right)
 \left(\begin{array}{cccccc}
                  &&&&& \\
                   &&&&& \\
                   &&c_{i,j}&&& \\
                   &&&&& \\
                   &&&&& \\
                   &&&&& \\
                   \end{array}
        \right)
\right|^{1/2}
\label{ProN2}
\end{eqnarray}
By choosing  $N$ consecutive sites for $x_1 \leq x_2 \leq .. x_N$,
the expressions (\ref{ProN},\ref{ProN2}) give the probability $Q(N)$ that $N$
consecutive  spins are parallel and the density $p_1(l)$ of domains of length
$l$ follows from a relation similar to (\ref{p1})
$$ Q(N) = \sum_{l \geq N} (l-N +1 )  p_1(l) \ . $$
\\ {\bf The long time regime} \\
In the long time limit, $c_{i,j}$  varies slowly with the positions
$x_i$ and $x_j$ of the  sites,  and becomes a continuous function
$c(x_i,x_j)$ of these two positions. Moreover,  for large $N$, all the sums in
the expression (\ref{ProN}) become integrals.
As in \cite{DHP1}, one can show that in this continuum limit,
(\ref{ProN}) or (\ref{ProN2}) can be rewritten as
\begin{equation}
Q(N)= \left[ \sqrt{1 - \mu \tilde{c}(N,N)}
 - \lambda \sqrt{ - \mu \tilde{c}(N,N)} \right]
 \ \exp \left[{1 \over 2} \makebox{tr} \log M \right] \
\label{QN}
\end{equation}
where this time
$c(x,y)=f({y-x \over  \sqrt{8t}})$ as in (\ref{cijbis},\ref{deff}),
the matrix $M$  is defined by
$$M(x,y) = \delta(x-y) + 2 \mu \  {d \over dx} c(x,y) \ ,$$
and the quantities appearing in (\ref{QN}) are given by
\begin{equation}
\makebox{tr} \log M =  -
 \sum_{n=1}^\infty {(-2 \mu)^{n} \over n} \int_{0}^{N} d x_1
\cdots \int_{0}^{N} d x_n \  {d \over dx} c(x_1, x_2) \cdots
 {d \over dx} c(x_n, x_1)
\label{TR}
\end{equation}
and
\begin{equation}
\tilde{c}(N,N)  \equiv   \int_0^N  dy \  c(N,y) M^{-1}(y,N) =
 \sum_{n=1}^\infty {(-2 \mu)^{n} } \int_{0}^{N} d x_1
\cdots \int_{0}^{N} d x_n  \  c(N, x_1) \  {d \over dx} c(x_1, x_2) \cdots
 {d \over dx} c(x_n, N) \ .
\label{PREF}
\end{equation}

If one normalizes all the distances on the lattice to   make the average domain
size become 1, one  defines $x$ by
\begin{equation}
N=  \sqrt{2\pi t}  \ {q \over q-1} \  x
\label{Nx}
\end{equation}
and the distribution of domain sizes $g(x;q)$ is then given by
$$Q(N)= \int_x^\infty dy (y-x) g(y;q) $$
so that in the long time limit, one ends up with
\begin{equation}
 \int_x^\infty dy (y-x) g(y;q) =
 \left[ \sqrt{1 - \mu A_1(x)}
 - \lambda \sqrt{ - \mu A_1(x)} \right] e^{A_2(x)}
\label{gxq}
\end{equation}
where
\begin{equation}
A_1(z) =
 \sum_{n=1}^\infty {(-2 \mu)^{n} } \int_{0}^{z} d x_1
\cdots \int_{0}^{z} d x_n  \  \gamma(z, x_1) \  {d \over dx} \gamma(x_1, x_2)
\cdots
 {d \over dx} \gamma(x_n, z) \ .
\label{A1}
\end{equation}
and
\begin{equation}
A_2(z) =  -  {1 \over 2}
 \sum_{n=1}^\infty {(-2 \mu)^{n} \over n} \int_{0}^{z} d x_1
\cdots \int_{0}^{z} d x_n \  {d \over dx} \gamma(x_1, x_2) \cdots
 {d \over dx} \gamma(x_n, x_1)
\label{A2}
\end{equation}
with
\begin{equation}
\gamma(x,y) = F(y-x) = {2 \over \sqrt{\pi}} \int_0^{\sqrt{\pi} (y-x) q /2(q-1)}
e^{-u^2} du
\label{gamma}
\end{equation}
and
$$ {d \over dx} \gamma(x,y) = - {q \over q-1}  \exp \left[{-\pi (y-x)^2 q^2
\over 4 (q-1)^2} \right] $$
Differentiating (\ref{gxq}) twice with respect to $x$
leads to the exact expression of $g(x;q)$.
\section{Expansions for the distribution of domain sizes}

The  exact expression (\ref{gxq}) for $g(x;q)$  requires the calculation of
$A_1$ and $A_2$  which we did not succeed to do explicitly for
general $q$ and $x$.
 However, it can be used to expand to arbitrary orders in various limits:
small $x$, large $x$,  large $q$.

\subsection{small $x$}
The small $x$ expansion is straightforward. One just need to expand
$\gamma(x,y)$ in (\ref{gamma})  in powers of $y-x$
$$ \gamma(x,y) =  F(y-x)={q\over q-1} (y-x)-{\pi\over 12} \left({q\over q-1}
\right)^3 (y-x)^3 + {\pi^2 \over 160} \left( {q \over q-1} \right)^5 (y-x)^5
...$$
and to replace $\gamma$ by its expansion in (\ref{gxq},\ref{A1},\ref{A2}).
 Mathematica
performs this to high order, and we found for $g(x;q)$ up to order $x^{13}$
\begin{eqnarray*}
g(x;q)&=&{\pi\over 2}{q\over q-1} x
-{\pi^2 \over 8}{q^3\over (q-1)^3}x^3
+{\pi^2\over 24}{q^3\over(q-1)^4}x^4
+{\pi^3\over 64}{q^5\over (q-1)^5}x^5\\
&-&{\pi^3\over 120}{q^5\over (q-1)^6}x^6
-{\pi^4 \over 768}{q^7\over (q-1)^7}x^7
+{\pi^4\over 26880}{q^6(3+23 q)\over(q-1)^8}x^8\\
&+&{\pi^5\over  12288 }{q^9\over(q-1)^9}x^9
-{\pi^5 \over 483840}{q^8(11+29q)\over (q-1)^{10}}x^{10}
-{\pi^6 \over 245760}{q^{11}\over (q-1)^{11}}x^{11}\\
&+&{\pi^6\over 170311680}{q^{10}(429 + 547 q)\over (q-1)^{12}}x^{12}\\
&+&{\pi^{6}\over 1857945600}{q^{10}(64-64q+315 \pi q^3 )\over (q-1)^{13}}
x^{13}\\
&+&O(x^{14}) \ .
\end{eqnarray*}
This expression is clearly different   from the one obtained (\ref{ind1})
in the uncorrelated approximation
\cite{ALEMA}.
 Note that only the first term
in both expansions is identical.

\subsection{Large x}
For domains much larger than the average  size, we expect
that, for finite $q$,  $g(x;q)$
  decays   exponentially
\begin{equation}
g(x;q)\simeq \exp[-A(q)x+B(q)]  \ .
\label{gasympt}
\end{equation}
 This can be understood by
recalling that a large domain  of size $L$ at finite $q$
is created (\ref{fq}) by combining $n$ domains  at $q=\infty$ with $n$
typically
$L/\langle l \rangle_{q=\infty}$. The probability of this is exponentially
small
in $n$. If the domains at $q=\infty$
were not correlated, the constants $A(q)$ and $B(q)$ would be given by
(\ref{ind2},\ref{ind3}).
 The
existence of correlations between the lengths of domains (\ref{COR})  alters
the coefficients of the
decay, but not the exponential decay itself.

The large $x$  behavior of $g(x;q)$ is directly connected (\ref{gxq}) to the
large $N$ behavior of $Q(N)$
\begin{equation}
Q(N) = \int_x^\infty dy (y-x) g(y;q) \simeq
  \exp[-A(q)x+B(q) - 2 \log A(q)]  \  .
\label{QN1}
\end{equation}
Therefore, we need  the large $x$ behavior of $A_1(x)$ and $A_2(x)$ given
by (\ref{A1},\ref{A2}). It turns out that (as in \cite{DHP1}) the prefactor $
\sqrt{1 - \mu A_{1}(x)} - \lambda \sqrt{ - \mu A_{1}(x)} $
vanishes as $x \to \infty$ when $q>2$ but has a  non zero finite limit when $q
< 2$.
One has therefore to examine these two cases separately.
\begin{itemize}
\item For $q<2$, the prefactor (which has been calculated in equation (35) of
\cite{DHP1}) has the following limit
\begin{equation}
  \sqrt{1 - \mu A_{1}(x)}
 - \lambda \sqrt{ - \mu A_{1}(x)} \rightarrow \sqrt{q(2-q)}
  \ \ \ \makebox{  for  } \ \ \
 x  \rightarrow \infty \ \  .
\label{pref}
\end{equation}
To evaluate  the large $x$ behavior of $A_2(x)$, one can use well known results
on Toeplitz determinants \cite{kac,kac1,kac2},
namely that for an even function $a(u)$ which decays as $u \to \pm \infty$
\begin{eqnarray}
 &&- \sum_{n=1}^\infty {(-2 \mu)^n \over n}
 \int_0^x du_1 \cdots \int_0^x du_n \ a(u_1-u_2) \cdots a(u_n-u_1)
\simeq
\nonumber \\
&&{x \over 2 \pi} \int_{-\infty}^\infty \log[ 1 + 2 \mu \tilde{a}(k) ] dk
 + \int_0^\infty  \ u \  du {1 \over 4 \pi^2}
\left|  \int_{-\infty}^\infty  e^{-iku} \log [ 1 + 2 \mu \tilde{a}(k) ] dk
\right|^2
\label{toeplitz}
\end{eqnarray}
 where
$$\tilde{a}(k)= \int_{-\infty}^\infty e^{iku} a(u) du $$
When this is used  in the case of $A_2(x)$ given by (\ref{A2}) and
(\ref{gamma}), one finds that
$$a(u)= -{q \over q-1} \exp \left\{-\pi u^2 q^2 \over 4(q-1)^2 \right\} $$
and
\begin{equation}
\tilde{a}(k)= -2  \exp \left\{- k^2 (q-1)^2 \over \pi q^2 \right\}
\label{ak}
\end{equation}
This leads to the following expressions in terms of $\mu = {q-1 \over q^2}$
\begin{equation}
A(q)= {1 \over 4} \ {q \over q-1} \sum_{n=1}^\infty {(4 \mu)^n \over n^{3/2}}
\label{resA}
\end{equation}
\begin{equation}
B(q)- 2 \log A(q) = {1\over 2} \log q(2-q) + {1 \over 4 \pi} \sum_{n=2}^\infty
{(4 \mu)^n \over n} \sum_{p=1}^{n-1} {1 \over \sqrt{p(n-p)}}
\label{resBB}
\end{equation}
which can be rewritten in order to make each term have a finite limit in the
previous expression  when
$q \to 2$
\begin{equation}
B(q)- 2 \log A(q) = \log q  - {q-1 \over q^2}  + {1 \over 4 \pi}
\sum_{n=2}^\infty {(4 \mu)^n \over n}   \left\{- \pi + \sum_{p=1}^{n-1} {1
\over \sqrt{p(n-p)}} \right\}  \  \ .
\label{resB}
\end{equation}

In (\ref{resA}) and (\ref{resB}), the series are convergent
for all values of $q$  (since \hbox{ $(q-1)/q^2 = \mu \leq 1/4$}) and so these
expressions can be  used to calculate numerically
$A(q)$ and $B(q)$. In section 5, we will compare the values predicted by
(\ref{resA})  and (\ref{resB}) to those (\ref{ind2},\ref{ind3}) of the
approximation where
the domains are uncorrelated.
\\ \\ {\bf Remark 1:} the radius of convergence of the sums which appear in
(\ref{resA},\ref{resB}) is $\mu=1/4$.  So as $\mu \to 1/4$, that is $q \to 2$,
these expressions could become singular.  As $\mu<1/4$  when $q>2$,
(\ref{resA}) and
(\ref{resB}) can be computed for  both for $q<2$ and $q>2$. However, it is easy
to see that these expressions as a function of $q$ are not analytic at $q=2$
(for example it is possible to show that
$ {d \over dq} {q-1 \over q} A(q) \to \sqrt{\pi}/4$  as $q \to 2$  so that
$A(q)$ given by (\ref{resA}) has a cusp at $q=2$). We will argue below, that
for $q>2$,  expressions (\ref{resA}) and (\ref{resB}) are no longer valid as
they are but should be replaced by their analytic continuation from the range
$q<2$.
\\ \\ {\bf Remark 2:}  a priori, in the Potts model, $q$ is an integer and so
the case $q<2$  is of little interest. However, Fortuin and Kasteleyn
\cite{FK}
have shown a long time ago that non integer values of $q$  have physical
realizations
as cluster models.
Here, it is very easy to check, that if one considers the zero temperature
dynamics of an Ising chain where the spins are initially uncorrelated but a
non zero magnetization $m$, the distribution of the  sizes  of  domains of $+$
spins is exactly the same as for the $q$ state Potts model when
$$q = {2 \over 1+m} \ .$$ In particular,  $q \to \infty$ corresponds to an
initial
condition where the $+$ spins are very rare whereas $q \to 1$ corresponds to
very few $-$ spins. So $m$ can take any value between $-1$ and $1$ and $q$ can
vary continuously
between 1 and $\infty$.

Also, for $q>2$, the reaction diffusion model (\ref{reactdif}) makes sense for
a
non-integer $q$.
\item
For  $q>2$, the large $x$ behavior of $A_2(x)$ is exactly the same as for
$q<2$.
However the prefactor in (\ref{gxq}) vanishes as $x \to \infty$.  The
evaluation
of this prefactor for large $x$ is not simple. A similar calculation was done
in
\cite{DHP1} and the result for $q>2$ turned out to be simply the analytic
continuation of the result obtained for $q<2$. Here we assume that this
property remains true for $g(x;q)$ (in the small $x$ expansion, at least,  one
can easily check that all the coefficients can be analytically continued at
$q=2$)
and that $A(q)$ and $B(q)$ in the range $q>2$ are given by the analytic
continuation $\overline{A(q)}    ,\overline{B(q)}$
of their expressions (\ref{resA},\ref{resB}) for $q<2$.
In the appendix, we give a way of deriving the following expressions of
$\overline{A(q)}$ and    $\overline{B(q)}$
\begin{equation}
\overline{A(q)} =
  {q \sqrt{\pi} \over q-1} \sqrt{- \log 4 \mu}  + {1 \over 4} {q \over q-1}
\sum_{n=1}^\infty {(4 \mu)^n \over n^{3/2} }
\label{Abar}
\end{equation}
\begin{eqnarray}
\overline{B(q)} - 2 \log \overline{A(q)} =
{1 \over 2} \log q (q-2)  -\log 4 - \log(-\log 4 \mu) + {1 \over 4 \pi}
\sum_{n=2}^\infty {(4 \mu)^n \over n} \sum_{p=1}^{n-1} {1 \over \sqrt{p(n-p)}}
\nonumber \\
-2 \sum_{n=1}^\infty {1 \over  n \sqrt{\pi}} \int_{\sqrt{-n \log 4 \mu}}^\infty
\ dv \ e^{-v^2}  \ .
\label{Bbar}
\end{eqnarray}

\end{itemize}
\subsection{  The large $q$ expansion }
In  the large $q$ limit, $\mu$, $\lambda$ and $\gamma(x,y)$ can be expanded in
powers of $1/q$ (see (\ref{LAMBDA},\ref{MU},\ref{gamma})).
 This leads to
$$ A_1(z)= - \mu  \ \gamma^2(0,z) + 4 \mu^2 \gamma(0,z) \  \int_0^z dx \
\gamma(0,x) \  { d  \gamma(x,z) \over dx} + O \left(  {1 \over q^3} \right) $$
$$ A_2(z)= - {x \over q} + O \left(  {1 \over q^2} \right) $$
and  to
\begin{eqnarray}
&&Q(N)=1-\int_{0}^{x}du e^{-\pi u^2/4}
\nonumber\\
&&+ {1\over q} \left\{-x(1-\int_{0}^{x}du e^{-\pi u^2/4})+
2\int_{0}^{x}du e^{-\pi u^2/4}-  x e^{-\pi x^2/4}-
2\int_{0}^{x}dy\int_{0}^{y}dz e^{-{\pi\over 4}(z^2+(x-y)^2)} \right\}
\nonumber\\
&&+O\left({1\over q^2}\right).
\end{eqnarray}
which becomes using (\ref{gxq})
\begin{equation}
g(x;q)= {\pi \over 2} x e^{- x^2 \pi /4} +
{1 \over q}  \left\{ - {\pi^2 \over 4} x^3 e^{-x^2 \pi /4}
 - {\pi \over 2} x^2 e^{-x^2 \pi /4}
 + {\pi \over 2} x e^{-x^2 \pi /4}
 + {\pi \over 2} x \int_0^x dz e^{-[(x-z)^2+z^2] \pi /4}
 \right\}
+O\left({1\over q^2}\right).
\label{g1ovq}
\end{equation}
This gives for the moments of the distribution  of the distribution $g(x;q)$
\begin{eqnarray}
&&\langle x^n\rangle=
\left( {2 \over \sqrt{\pi}} \right)^n \left[
  \Gamma\left( {n \over 2} +1 \right) \right.  \nonumber \\
&&  \left. + {1 \over q} \left\{
 - (n+1) \Gamma\left( {n \over 2} +1 \right)
   - {2 \over \sqrt{\pi}} \Gamma\left( {n+1 \over 2} +1 \right)
+ {2 \over \sqrt{\pi}}
\sum_{p=0}^n \left(\begin{array}{c}
n \\ p \end{array} \right)   \Gamma\left( {p+1 \over 2} \right)
\Gamma\left( {n-p+2 \over 2} \right)
\right\} \right]  \nonumber \\
&&+O\left({1\over q^2}\right).
\end{eqnarray}
 and in particular,
$$\langle x^2 \rangle = {4 \over \pi} \left(
1 +  {1 \over 2 q} \right)
+O\left({1\over q^2}\right). $$
\\ \\ {\bf Remark:}  because of (\ref{fq}) all the results to order $1/q$ can
be
recovered from the knowledge (\ref{ginf},\ref{g2inf}) of $g(x;\infty)$ and
$g_2(x,x';\infty)$ and it is easy to check that
$$g(x;q) = \left( 1 + {1 \over q} \right) \left[ \left( 1 - {1 \over q} \right)
 g \left( ( 1 + {1 \over q} ) x ; \infty \right) + { 1 \over q} \int_0^{(1 + {1
\over q}) x} dz \
g_2 \left( ( 1 + {1 \over q} ) x -z, z ; \infty \right)  \right]
+O\left({1\over q^2}\right)$$
is equivalent to (\ref{g1ovq}).
It is straightforward in principle to generate higher order terms
in ${1\over q}$ but the expressions become quite complicated.

\section{Simulations}
In order to check the validity of our results against the predictions of the
independent interval approximation \cite{ALEMA}, we made a simulation.
Because of ({\ref{fq}),  the distribution of  domain sizes for all
values of $q$ can be extracted from a single simulation done at $q= \infty$.
One starts with a  random initial condition at $q=\infty$ (this is done by
choosing initially each spin  with a different value: for example spin $i$ is
given color $i$)   and let the system coarsen according to  zero temperature
Glauber dynamics.
Once  the  system at $q=\infty$ has evolved for a certain time, one obtains  a
system for an arbitrary value (integer or non-integer) of $q$ by removing each
domain wall with  probability $1/q$.

The easiest properties one can measure are the moments of the distribution of
domain sizes. Our data
(for a system of $10^6$ spins)
for the second moment $\langle x^2 \rangle$   of the distribution $g(x;q)$ of
domain sizes for $q=2, 3 $ and $5$  are shown in figure 1  as a function of the
average size $\langle l \rangle_\infty$ of the domains for $q=\infty$.
 For Glauber dynamics,  this average  $\langle l \rangle_\infty$ increases with
time like $t^{1/2}$. The larger $\langle l \rangle_\infty$ is, the closer we
are from the asymptotic regime. However, as the simulation is done for a
finite system, the results become  noisy if one waits too long simply because
the number of domains is
too small to allow good statistics.

These numerical results can be compared to the approximation where the
correlations are neglected (\ref{indep})
\begin{equation}
\langle x^2 \rangle_{\makebox{independent}} = {q-1 \over q} {4 \over \pi} + {2
\over q}  \ .
\label{ind7}
\end{equation}
We see in figure 1 that our numerical data seem stable and accurate enough to
show a clear discrepancy with the approximation.
Unfortunately, our exact expression (\ref{gxq}) is complicated and we were
unable to obtain closed expressions of the moments for the whole range of $q$.
We could test however our conviction that correlations between the domains at
$q=\infty$ have an effect.
If in the calculation of $\langle x^2 \rangle$, we neglect the correlations
between all pairs of domains except for nearest neighbor domains (\ref{COR}),
we  find an improved estimate
\begin{equation}
\langle x^2 \rangle_{\makebox{improved}} = {q-1 \over q} {4 \over \pi} + {2
\over q}  + 2 { q-1 \over q^2} \left( {3 \over \pi} -1 \right)
\label{impro}
\end{equation}
which, although not exact,  seems to be in much better agreement with the
results of the simulations.

As we could not calculate the moments from (\ref{gxq}),
we  tried to test our results by measuring the exponential tail of the
distribution $g(x;q)$. It is always difficult to measure numerically a
probability distribution because one has to use bins: if they are too narrow,
the data are noisy and if they are too broad, all the details of the
distribution are smoothened out.
To avoid this difficulty, we measured the integrated distribution
$h(x;q)$
$$h(x;q)= \int_x^\infty g(y;q) dy$$ and we compared it to our prediction
(\ref{gasympt}) for the tail    (\ref{resA},\ref{resB}) for $q<2$ and
(\ref{Abar},\ref{Bbar}) for $q>2$. We also compared the results of our
simulations to
the prediction of the independent domains approximation
(\ref{ind2},\ref{ind3}).
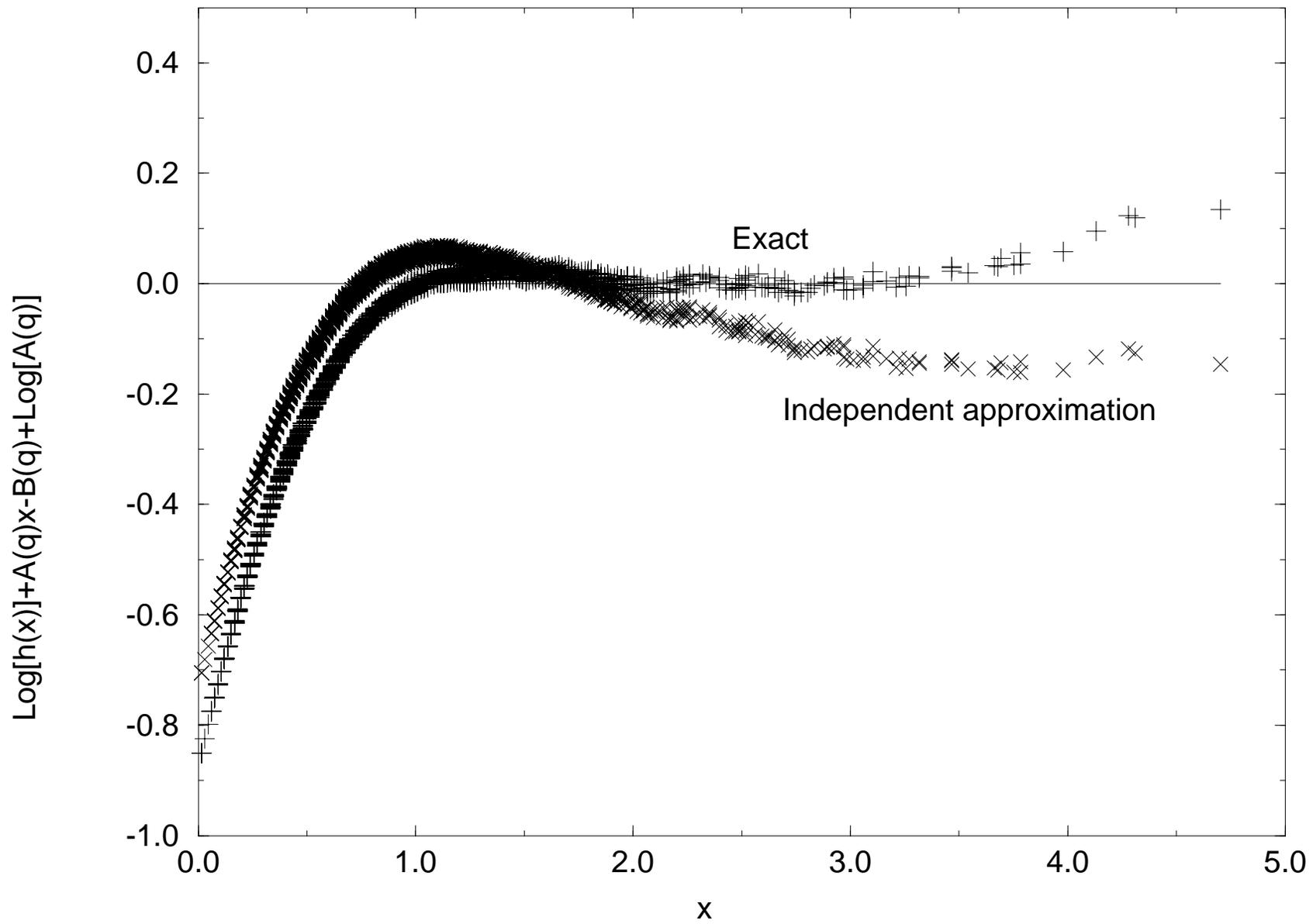
In figure 2, we plot the logarithm of the integrated distribution divided by
its predicted form
 $[A(q)x  -B(q) + \log A(q)] + \log h(x;q) $ versus $x$ for $q=5$ for a system
of 2 millions spins,  after $1000$ updatings per spin. We see  that the
agreement is very good when (\ref{Abar},\ref{Bbar}) are used whereas for
(\ref{ind2},\ref{ind3}), there is a small but visible discrepancy. Of course,
for $x$ too large, our data are too noisy  and for $x$ too small,
 the asymptotic form has no reason to be valid , so that the range where the
agreement is the best is $1 < x < 3.5$.

In table 1,  the exact  values of $A(q)$ and $B(q)$ obtained from
(\ref{resA},\ref{resB}) for $q<2$ and
(\ref{Abar},\ref{Bbar}) for $q>2$  are compared to those coming from the
independent domains approximation  (\ref{ind2} and \ref{ind3}). We see that,
although different,  they are very close.

\section{The reaction diffusion model}
The relation between the spin dynamics and the reaction diffusion model
(\ref{reactdif}) can be used to  calculate all kinds of properties of the
reaction diffusion model.
For the zero temperature Glauber dynamics  of the Potts chain, it results from
 the  analogy to random walks (section 2) that
$$ \makebox{Probability} \{
S_{x_1}(t)\neq S_{x_2}(t) \}
=\left({q-1 \over q} \right)   c_{1,2} $$
and that (\ref{1234a}) for $x_1<x_2<x_3<x_4$
$$  \makebox{Probability} \{
S_{x_1}(t)\neq S_{x_2}(t) \makebox{ and } S_{x_3}(t)  \neq S_{x_4}(t) \}
=\left({q-1 \over q} \right)^2  \left[ c_{1,3} + c_{2,4} - c_{2,3} - c_{1,4} +
c^{(2)}_{1,2,3,4}  \right]  \ .$$
In the long time regime, this becomes using (\ref{cijbis}) and (\ref{deff})
for $x_1=0, x_2=1, x_3=r,x_4=r+1$
$$ \makebox{Probability} \{
S_{0}(t)\neq S_{1}(t) \} \simeq  H'(0) $$
$$  \makebox{Probability} \{
S_{0}(t)\neq S_{1}(t) \makebox{ and } S_{r}(t)  \neq S_{r+1}(t) \}
 \simeq \left({q-1 \over q} \right)^2  \left[-H''(r) + H''(r) H(r) - (H'(r))^2
+ (H'(0))^2  \right] $$
where $$H(r)={2 \over \sqrt{\pi}} \int_0^{r/\sqrt{8t}} e^{-u^2} du $$

If these expressions are interpreted in terms of  domain walls (i.e. of
particles in the reaction  diffusion model), one obtains
that the density $\rho$ of particles and the probability $\rho_2(0,r)$ of
finding a pair of particles at $0$ and at $r$
are given by
$$  \rho
\simeq  {q-1 \over q} H'(0) = {q-1 \over q}
{1 \over \sqrt{2 \pi t }} $$
$$  \rho_2(0,r)
 \simeq \left({q-1 \over q} \right)^2  \left[-H''(r) + H''(r) H(r) - (H'(r))^2
+ (H'(0))^2  \right]  \  .$$

By eliminating the time $t$ between these two equations one finds that
in the long time limit, the probability of finding a particle at $0$
and $r$ is given by
\begin{equation}
\rho_2(0,r) = \rho^2
\left [ 1- e^{- 2 z^2} + 2 z  e^{-z^2}
\int_z^\infty e^{-u^2} du \right]
\label{scaling}
\end{equation}
and
$$z= {q \over q-1} {\sqrt{\pi} \over 2}  \ \rho \ r  $$
We see in particular that the correlations decay  like a Gaussian
instead of the "usual" exponential. \\ \\
{\bf Remark:} it has already been noticed \cite{Bur1,Bur2,Bur3,Bur4,Bur5} that
several properties of reaction diffusion models can be calculated exactly, by
writing closed equations for
the probability that an interval of length $l$ is empty (or two intervals are
empty...). This method  (which is  rather close in spirit to our approach as it
essentially considers that walkers which do not meet in
one dimension are free fermions) could also be used to  recover
(\ref{scaling}).

\section{Conclusion}
The long time regime of domain growth phenomena is by many respects analogous
to
what happens near a critical point in a second order phase transition:
this regime is characterized both by universal exponents and by universal
scaling functions.
In the present paper, we have determined the distribution of domain sizes in
this regime, for the $q$-state Potts model in one dimension.
This distribution is universal, in the sense that,  at least,
it would remain the same for short ranged
 correlations in the initial condition.

The exact expression we found is rather complicated (\ref{gxq}-\ref{gamma}) but
it can be simplified in several limits (section 4). Our exact results are
different
(although rather close to)  the predictions of a simple approximation
proposed recently \cite{ALEMA}. However, with sufficient numerical effort, we
think that our
simulations of section 5  indicate the  validity of our results against those
(\ref{ind2}-\ref{ind3})
of that approximation.

The present work  has a lot in common with \cite{DHP1} where the exponent
characteristic of the persistent spins have been calculated exactly
for the same system.   Here we are concerned with space properties
whereas   \cite{DHP1} was devoted to time properties.
The existence of non trivial exponents for the number of persistent spins
\cite{DBG,kra,Stau,krap,DOS}
in $d>1$  has certainly its counterpart in terms of distribution of domain
sizes.
However, in $d>1$, the size is one among many ways of characterizing a domain
 (its volume, its perimeter, its shape
\cite{rut}, its number of neighboring domains...). It would be interesting to
know
whether the probability distributions of all these characteristics of domains
are universal  in $d>1$.
Another open question  for systems in $d>1$ is how to define the domains in
presence of thermal noise (i.e. at non-zero temperature) and how to distinguish
them from thermal fluctuations. When the typical size of domains is much larger
than the equilibrium correlation length,  it seems intuitively  easy to make
this distinction, but to our knowledge, a clean way of measuring the size of
domains (in presence of noise) has not yet been proposed. \\ \\ \\ \\
We thank Vincent Hakim and Vincent Pasquier for useful discussions. \\ \\ \\ \\

\section{Appendix : Analytic Continuations}
\setcounter{equation}{0}
\def\theequation{A\arabic{equation}}

In this appendix we  show how to analytically continue the expressions
(\ref{resA}, \ref{resB}) of $A(q)$ and $B(q)$
obtained in the large $x$ expansion for $q<2$  to the  range $q>2$.
It is easier to start with the expressions coming from
(\ref{QN1}-\ref{toeplitz}), that is
\begin{equation}
A(q)={-1\over 4\pi}\int_{-\infty}^{\infty}dk\log[1+2\mu \tilde{a}(k)]\\
\label{Aq1}
\end{equation}
\begin{equation}
B(q) - 2 \log A(q)={1\over 2}\log q(2-q)+
{1\over 8\pi^2}\int_{0}^{\infty} y  \ dy
 \left|\int_{-\infty}^{\infty} dk e^{iky}
\log[1+2\mu \tilde{a}(k)]\right|^2
\label{Bq1}
\end{equation}
where $\tilde{a}(k)$ is given by (\ref{ak})
\begin{equation}
\tilde{a}(k)= -2  \exp \left\{- k^2 (q-1)^2 \over \pi q^2 \right\}
\label{ak1}
\end{equation}
As $q \to 2$, that is $\mu \to 1/4$, the integrands in (\ref{Aq1}) and
(\ref{Bq1})
 become singular at $k=0$:   two complex zeroes in the  complex plane of the
variable $k$
approach $k=0$ as $q \to 2$ and  the exchange of these two  zeroes is at the
origin of the analytic continuation.
\\ \\
{\bf A generic example:} to make the analytic continuation of
(\ref{Aq1}) and (\ref{Bq1}) simpler to understand
we treat first a case where the expression (\ref{ak1}) of
$\tilde{a}(k)$ would be replaced by
\begin{equation}
\tilde{a}(k)= {- 2  \over 1 + Q(k)}
\label{aQ}
\end{equation}
 where $Q(k)$ is a real polynomial of degree $2n$ with ($Q(k) \sim k^2$ as $k
\to 0$ and such that $Q(k)+1$ and $Q(k) +1 - 4 \mu$ have both  only complex
zeroes).
One can therefore write
\begin{equation}
1 + 2 \mu \tilde{a}(k) = 1 - {4 \mu \over 1 + Q(k)} = \prod_{j=1}^n {(z_j -
k)(z_j^* - k) \over
 (y_j - k)(y_j^* - k) }
\label{frac}
\end{equation}
where the $y_j$ are the $n$  zeroes of $1 + Q(k)$ with positive imaginary parts
and the $z_j$ are the $n$ zeroes of
$1 - 4 \mu + Q(k)$ with positive imaginary parts.
One can rewrite (\ref{Aq1},\ref{Bq1}) as
\begin{equation}
A(q)={1\over 4\pi}\int_{-\infty}^{\infty}dk {2   \mu   \ k  \ \tilde{a}'(k)
\over 1 + 2 \mu \tilde{a}(k)}
={1\over 4\pi}\int_{-\infty}^{\infty}dk
{4\mu \ k \  Q'(k)\over [1+Q(k)][1-4 \mu +Q(k)]}
\label{Aq2}
\end{equation}
\begin{equation}
B(q) -2 \log A(q)=
  {1\over 2}\log q(2-q) +
{1\over 8\pi^2}\int_{0}^{\infty} { dy \over y} \left|\int_{-\infty}^{\infty} dk
\  e^{iky}
{4\mu Q'(k)\over [1+Q(k)][1-4 \mu +Q(k)]} \right|^2
\label{Bq2}
\end{equation}
and find using the theorem of residues
\begin{equation}
A(q)={i \over 2} \sum_{j=1}^n z_j - y_j
\label{Aq3}
\end{equation}
and
\begin{equation}
B(q) - 2 \log A(q) = {1 \over 2} \log q ( 2-q)
+ {1 \over 2} \sum_{j=1}^n \sum_{j'=1}^n \log \left\{
 \left( {i y_{j'}^* - i z_j \over i y_{j'}^* - i y_j } \right)
 \left( {i z_{j'}^* - i y_j \over i z_{j'}^* - i z_j } \right)  \right\}
\label{Bq3}
\end{equation}
These expressions (\ref{Aq3}) and (\ref{Bq3}) are valid for $q<2$. To obtain
their analytic continuation $\overline{A(q)},\overline{B(q)}$   to  the range
$q>2$, we   have essentially to exchange the role of $z_1$ and $z_1^*$, the two
zeroes of $1-4 \mu +Q(k)$ which go through $k=0$ when
$q=2$.  So for $q>2$,
\begin{equation}
\overline{A(q)}=  {i \over 2} (z_1^* - z_1) + {i \over 2} \sum_{j=1}^n z_j -
y_j
\label{Aq4}
\end{equation}
and
\begin{eqnarray}
\overline{B(q)} - 2 \log \overline{A(q)} =
\left[{1 \over 2} \log q (q-2) - i {\pi \over 2} \right] +
\left[{1 \over 2} \log \left| {i z_1^* - i z_1 \over i z_1 - i z_1^*} \right| +
i {\pi \over 2} \right] +
\nonumber \\
 {1 \over 2} \sum_{j=1}^n  \log \left\{
 \left( {i y_{j}^* - i z_1^* \over i y_{j}^* - i z_1 } \right)
 \left( {i z_{1} - i y_j \over i z_{1}^* - i y_j } \right)  \right\}
 + {1 \over 2} \sum_{j=2}^N  \log \left\{
 \left( {i z_{j}^* - i z_1 \over i z_{j}^* - i z_1^* } \right)
  \left( {i z_{1}^* - i z_j \over i z_{1} - i z_j } \right)  \right\}
\nonumber \\
+ {1 \over 2} \sum_{j=1}^n \sum_{j'=1}^n \log \left\{
 \left( {i y_{j'}^* - i z_j \over i y_{j'}^* - i y_j } \right)
 \left( {i z_{j'}^* - i y_j \over i z_{j'}^* - i z_j } \right)  \right\}
\label{Bq4}
\end{eqnarray}
where we have added an infinitesimal imaginary part to $q$ to define
the analytic continuation of $\log(iz_1-iz_1^*)$ and $\log q(2-q)$.

Using the fact that
$$\prod_{j=1}^n (z_1- y_j)(z_1- y_j^*) = 1 + Q(z_1) = 4 \mu$$
and that
$$\prod_{j=2}^n (z_1- z_j)(z_1- z_j^*) = { Q'(z_1) \over z_1 - z_1^*}$$
(\ref{Bq4}) becomes
\begin{eqnarray}
\overline{B(q)} - 2 \log \overline{A(q)} =
{1 \over 2} \log q (q-2)
+\log 4 \mu  - \log | Q'(z_1)| + \log | z_1 - z_1^*|
\nonumber \\
- \sum_{j=1}^n  \log \left\{
 (z_1^* - y_j)(z_1- y_j^*)  \right\}
 +  \sum_{j=2}^N  \log \left\{
  (z_1 - z_j^*)(z_1^* - z_j)  \right\}
\nonumber \\
+ {1 \over 2} \sum_{j=1}^n \sum_{j'=1}^n \log \left\{
 \left( {i y_{j'}^* - i z_j \over i y_{j'}^* - i y_j } \right)
 \left( {i z_{j'}^* - i y_j \over i z_{j'}^* - i z_j } \right)  \right\}
\end{eqnarray}
This can be further transformed using the following identity
$$
 {1 \over 2 \pi i} \int_{-\infty}^\infty dk \left( {1 \over k - z_1} -
 {1 \over k - z_1^*} \right) \log \left( { (k-z_j)(k-z_j^*) \over (k-y_j)
(k-y_j^*)}\right) =
\log \left( { (z_1-z_j^*)(z_1^* -z_j) \over (z_1-y_j^*)(z_1^*-y_j)}\right)
$$
to become
\begin{eqnarray}
\overline{B(q)} - 2 \log \overline{A(q)} =
{1 \over 2} \log q (q-2)
+\log 4 \mu  - \log | Q'(z_1)| - \log | z_1 - z_1^*|
\nonumber \\
+{1 \over 2 \pi i} \int_{-\infty}^\infty dk \left( {1 \over k - z_1} - {1 \over
k - z_1^*} \right) \log \left( 1 - {4 \mu \over 1 + Q(k)} \right)
\nonumber \\
+ {1 \over 2} \sum_{j=1}^n \sum_{j'=1}^n \log \left\{
 \left( {i y_{j'}^* - i z_j \over i y_{j'}^* - i y_j } \right)
 \left( {i z_{j'}^* - i y_j \over i z_{j'}^* - i z_j } \right)  \right\}
\end{eqnarray}
Lastly, using the identity
$$ 2 \mu \tilde{a}'(z_1) = {4 \mu Q'(z_1) \over [1 + Q(z_1)]^2 } = {Q'(z_1)
\over 4 \mu} $$
one finds for $q >2$
\begin{eqnarray}
\overline{B(q)} - 2 \log \overline{A(q)} =
{1 \over 2} \log q (q-2)
  - \log | 2 \mu  \tilde{a}'(z_1)| - \log | z_1 - z_1^*|
\nonumber \\
+{1 \over 2 \pi i} \int_{-\infty}^\infty dk \left( {1 \over k - z_1} - {1 \over
k - z_1^*} \right) \log \left( 1 + 2 \mu \tilde{a}(k)  \right)
\nonumber \\
+{1\over 8\pi^2}\int_{0}^{\infty} y dy
 \left|\int_{-\infty}^{\infty} dk e^{iky}
\log[1+2\mu \tilde{a}(k)]\right|^2
\label{Bq5}
\end{eqnarray}
This expression together with (\ref{Aq4})
\begin{equation}
\overline{A(q)} = {i \over 2} (z_1^* - z_1) - {1 \over 4 \pi}
\int_{-\infty}^\infty dk \log[1 + 2 \mu \tilde{a}(k) ]
\label{Aq5}
\end{equation}
give for the generic example (\ref{ak1}) the analytic continuation
$\overline{A(q)},
\overline{B(q)}$ of $A(q)$ and $B(q)$ to the range $q>2$.

Although (\ref{Bq5},\ref{Aq5}) were derived for $\tilde{a}(k)$ of the form
(\ref{ak1}), it should remain valid for a broader class of functions,
probably as long as $\tilde{a}(k)$ is such that only a pair of zeroes of $1 + 2
\mu
\tilde{a}(k)$ exchange as $q$ crosses 2.

We believe that (\ref{Bq5}) and (\ref{Aq5}) remain valid for
$\tilde{a}(k)$ given by (\ref{ak})
so that
$$z_1 = i  \  {q  \ \sqrt{\pi} \over q-1} \sqrt{- \log 4 \mu} $$
\begin{equation}
\overline{A(q)} =
  {q \sqrt{\pi} \over q-1} \sqrt{- \log 4 \mu}  + {1 \over 4} {q \over q-1}
\sum_{n=1}^\infty {(4 \mu)^n \over n^{3/2} }
\label{Aq6}
\end{equation}
\begin{eqnarray}
\overline{B(q)} - 2 \log \overline{A(q)} =
{1 \over 2} \log q (q-2)  -\log 4 - \log(-\log 4 \mu) + {1 \over 4 \pi}
\sum_{n=2}^\infty {(4 \mu)^n \over n} \sum_{p=1}^{n-1} {1 \over \sqrt{p(n-p)}}
\nonumber \\
-2 \sum_{n=1}^\infty {1 \over  n \sqrt{\pi}} \int_{\sqrt{-n \log 4 \mu}}^\infty
\ dv \ e^{-v^2}
\label{Bq6}
\end{eqnarray}

\begin{table}[p]\label{Asymptotic decay}
\begin{tabular}{|l|c|c|c|c|c|} \hline
$q$ &Exact $A(q)$&Exact $B(q)$&Approximate $A(q)$&Approximate $B(q)$ \\
\hline
    1.2      & 1.0769&    .153     & 1.0675&    .133 \\
    1.5      & 1.1748&    .344     & 1.1532&    .299 \\
    2.       & 1.3062&    .597     & 1.2685&    .517\\
    3.       & 1.4998&    .963     & 1.4370&    .832 \\
    5.       & 1.7537&   1.439     & 1.6630&   1.239 \\
    10.      & 2.1066&   2.101     & 1.9777&   1.809\\
    100.     & 3.2273&   4.369     & 3.0065&   3.804\\
\hline
\end{tabular}
\caption{
 The exact  values of $A(q)$ and $B(q)$ obtained from
  (\ref{resA},
\ref{resB}) for $q<2$ and
(\ref{Abar},\ref{Bbar}) for $q>2$  are compared to those coming from the
independent domains approximation  (\ref{ind2},\ref{ind3}).}
\end{table}
{\noindent\LARGE Figure caption.}
\begin{description}
\item [Fig.~1.]
The second moment $\langle x^2 \rangle$ versus $\langle l \rangle_\infty$ for
$q=2$, $q=3$, $q=5$
for a system of $10^6$ spins. The dashed lines indicate the values predicted by
the improved approximation (\ref{impro}) and the plain  lines those of the
independent approximation (\ref{ind7}).
\item [Fig.~2.]
The logarithm of the (measured) integrated distribution $h(x;q)$ divided by its
predicted asymptotic  form $A(q) x - B(q)+ \log A(q) + \log h(x;q)$ for a
system of
$2 \times 10^6$ spins after $1000$ updatings per spin. When the exact
expressions of $A(q)$ and $B(q)$ are used, the  agreement is   very
satisfactory,
whereas when on takes the values predicted by the independent domain
approximation,
one sees a clear discrepancy.
\end  {description}

\newpage
\begin{thebibliography}{20}
\bibitem{Bray}  A.J. Bray, Adv. Phys. {\bf 43}, 357 (1994)
\bibitem{Glauber} R.J. Glauber, J. Mat. Phys. {\bf 4}, 294 (1963)
\bibitem{Br} A.J. Bray, J. Phys.\ A {\bf 23}, L67 (1990)
\bibitem{AF} J.G. Amar and F. Family, Phys.\ Rev.\ A {\bf 41}, 3258 (1990)
\bibitem{ABD}  D. ben-Avraham, M.A. Burschka and C.R. Doering,
J. Stat. Phys. {\bf 60}, 695 (1990)
\bibitem{DGY} B. Derrida, C. Godr\`{e}che and I. Yekutieli,
 Phys.\ Rev.\ A {\bf 44}, 6241 (1991)
\bibitem{DHP} B. Derrida, V. Hakim and V. Pasquier, Phys. Rev. Lett. {\bf 75},
751 (1995)
\bibitem{DHP1} B. Derrida, V. Hakim and V. Pasquier, preprint 1996,
 submitted to J. of Stat. Phys.
\bibitem{Ra} Z. R\'acz, Phys. Rev. Lett. {\bf 55}, 1707 (1985)
(1990)
\bibitem{Pr} V. Privman, J. Stat. Phys. {\bf 69}, 629 (1992)
\bibitem{voter} T.M. Liggett, {\it Interacting Particle Systems},
NY: Springer Verlag (1985)
\bibitem{Der} B. Derrida, J. Phys.\ A {\bf 28}, 1481  (1995)
\bibitem{dG1} P.G. De Gennes, J. Chem. Phys. {\bf 48}, 2257   (1968)
\bibitem{dG2}J. Villain and P. Bak, J. Physique {\bf 42}, 657 (1981)
\bibitem{dG3} M.E. Fisher, J. Stat. Phys.  {\bf 34},  667  (1984)
\bibitem{ALEMA} P.A. Alemany and D. ben-Avraham, Phys. Lett. \ A {\bf 206}, 18
(1995)
\bibitem{kac} G. Szeg\"o, Commun. S\'em. Math. Univ. Lund 228 (1952)
\bibitem{kac1} M. Kac, Duke Math. J {\bf 21}, 501 (1954)
\bibitem{kac2} B. M. McCoy and T.T. Wu, {\em The
Two-Dimensional Ising Model}, Harvard U. Press, Cambridge, Mass. (1973) and
references therein.
\bibitem{FK} F.C. Fortuin and P.W. Kasteleyn, Physica \ A {\bf 57}, 536 (1972)
\bibitem{Bur1}C.R. Doering, Physica A188, 386 (1992)
\bibitem{Bur2} M.A. Burschka, Europhys. Lett. 16, 537 (1991)
\bibitem{Bur3}
K. Krebs, M.P. Pfanmnm\"uller,  B. Wehefritz and H. Hinrichsen, J. Stat. Phys.
{\bf 78}, 1429 (1995)
\bibitem{Bur4} D. ben-Avraham, Mod. Phys. Lett. B9, 895 (1995)
\bibitem{Bur5}
M.A. Burschka and  C.R. Doering, to be published
\bibitem{DBG} B. Derrida, A. J. Bray and C. Godr\`{e}che,
  J. Phys.\ A {\bf 27}, L357 (1994)
\bibitem{kra} P.L. Krapivsky, E. Ben-Naim and S. Redner, Phys. Rev. E {\bf 50},
2474 (1994)
\bibitem{Stau} D. Stauffer,   J. Phys.\ A {\bf 27}, 5029 (1994)
\bibitem{krap} E. Ben-Naim, L. Frachebourg and P. L. Krapivsky,  preprint 95
\bibitem{DOS} B. Derrida, P.M.C. de Oliveira and D. Stauffer,    Physica A {\bf
224}, 604 (1996)
\bibitem{rut}  A.D. Rutenberg, preprint 1996



\end{thebibliography}
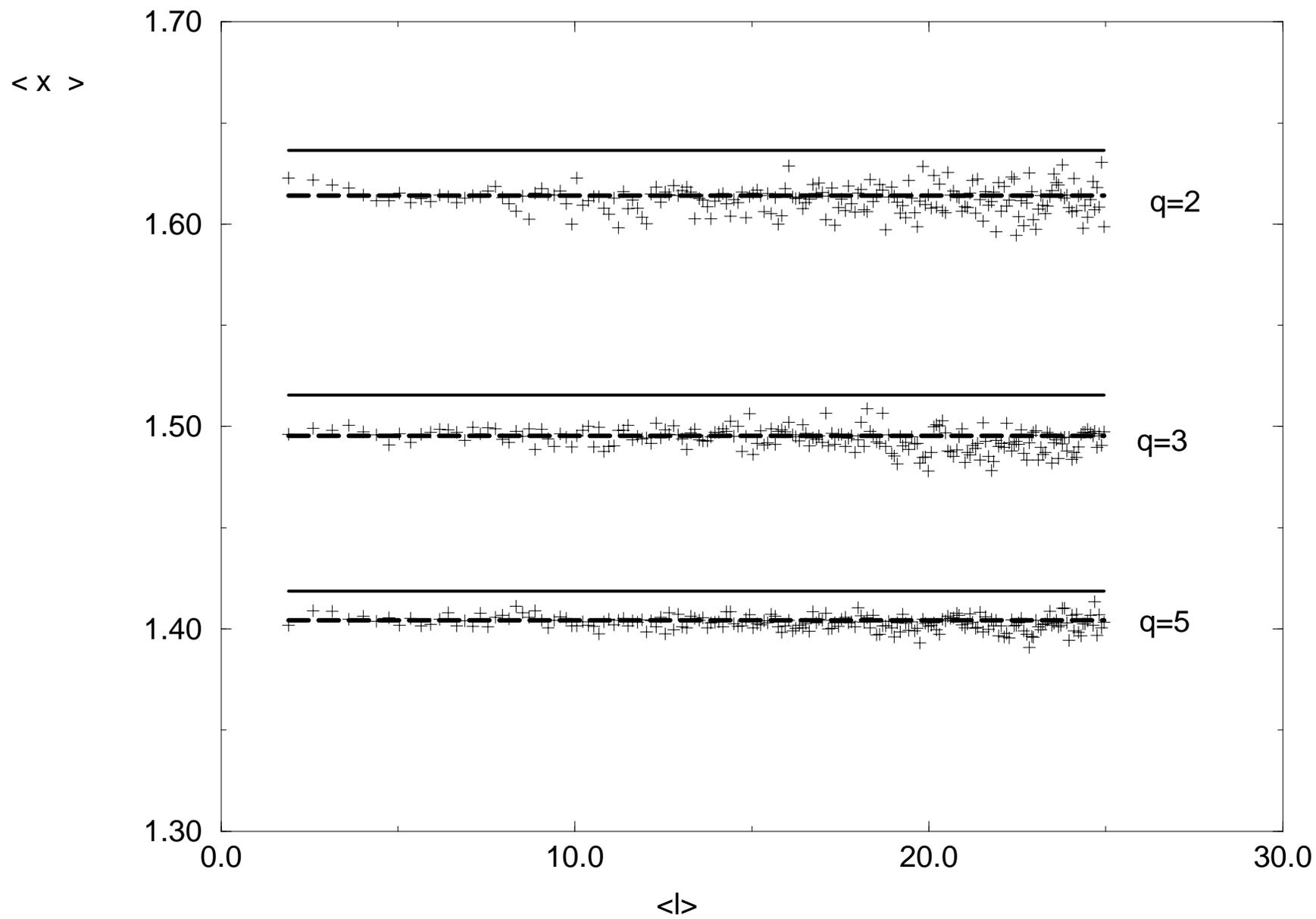
\end{document}